\documentstyle[aps]{revtex}
\begin{document}
\draft
\title{Evidence of structural instability near $T^{*}\sim 250$ K in the
resistivity of Bi$_2$Sr$_2$CaCu$_2$O$_8$ whiskers}
\author{Weimin Chen, J. P. Franck, and J. Jung}
\address{Department of Physics, University of Alberta, Edmonton T6G 2J1, Canada}
\date{\today}
\maketitle
\begin{abstract}

We report anomalous features in the normal-state resistivity of
single crystalline Bi$_2$Sr$_2$CaCu$_2$O$_8$ whiskers near
$T^{*}\sim 250$ K. From these varied oxygen-doping samples, the
resistance ($R$) was found to deviate {\it upward} from the
linear-$T$ relation for $T < T^{*}$, while $R$ keeps decreasing
with $T$. The deviation, $\Delta R \equiv R - (a + bT)$, reaches a
maximum in a temperature range of $\sim 25$ K, just before $R-T$
follows the well-known descent associated with the pseudogap
opening in underdoped samples. A second kink was also found in
some samples immediately following the first one, resulting in an
S-shaped feature in the $R-T$ curve. The resistance then resumes
nearly linear-$T$ decrease below the anomalies in contrast to the
first case. We interpret these features as being related to
crystal structure transformation (or lattice distortion), as was
evidenced in thermal and elastic property measurements. These
structural instabilities seem to be connected to the subsequent
dynamics as $T$ is further lowered.
\end{abstract}


   The normal state of high-$T_{c}$ cuprates provides rich
information for an understanding of superconducting mechanism. Two
of the many interesting normal-state properties are the so-called
pseudogap opening and crystal structural transformation. In
resistivity ($\rho$) measurement, the former was believed to
manifest itself as a downturn deviation from the linear $\rho-T$
relation below some characteristic temperature $T^{*}> T_{c}$ in
underdoped samples. Evidence for crystal structure transformation
(or lattice distortion) is more difficult to detect because of
such factors as crystal quality and the weakness of its physical
effects. Structural transformations are generally studied by
elastic and acoustic methods \cite{1}. In most of the cuprates,
anomalous features corresponding to structural instabilities in
the normal state have been observed. In La$_{2-x}$Sr$_x$CuO$_4$, a
tetragonal to orthorhombic structure transition upon cooling
occurred at 220 K \cite{2}. Around the same temperature for
YBa$_2$Cu$_3$O$_{7-\delta}$, the sound velocity shows a step-like
change and was associated with the instability corresponding to a
structural order-disorder transition \cite{3}. A step-like change
in Young's modulus (hysteresis) also appeared in elastic
measurements of single crystals at 200 - 240 K \cite{4,5}.  In Bi
cuprates, similar evidence was also accumulated. Dominec {\it et
al.} \cite{6} reported sound velocity deviations around 240 K in
Bi$_2$Sr$_2$CuO$_6$ (Bi-2201). Acoustic studies revealed sound
velocity softening along the $a$- axis around 250 K in
Bi$_2$Sr$_2$CaCu$_2$O$_8$ (Bi-2212) samples \cite{7,8}. In elastic
property measurements of similar Bi-2212 whiskers as studied in
the present work, Tritt {\it et al.} \cite{9} found anomalous
peaking of Young's modulus and stress-strain hysteresis near 270
K, and interpreted it as evidence for a phase transition. Similar
results were also obtained for Bi$_2$Sr$_2$Ca$_2$Cu$_3$O$_{10}$
(Bi-2223) \cite{10}. More directly, by using acoustic and x-ray
diffraction methods, Titova {\it et al.} \cite{11} consistently
found a small increase of lattice dimensions at $T \sim 190 - 270
$K for the cuprates: YBCO (123 and 124), Bi-2201, 2212, and 2223,
and Ti-series compounds.
   The structural instabilities of the cuprates observed at
comparable temperature ranges (about 200 to 300 K) demonstrate
common features consistent with their similar crystalline
structures. Obviously, structural transformation affects transport
properties of the material and should be detectable by resistivity
measurement. Indeed, resistivity jumps were observed above $T_{c}$
in La$_{2-x}$Sr$_x$CuO$_4$, which are caused by a structural phase
transition \cite{12}. However, few results were reported for other
cuprates in the temperature regions where lattice instabilities
were found by mechanical methods described above. This situation
can be understood in two ways. First, the influence of structural
instabilities on sample resistivity is small, because at such
structural transformation, the crystal symmetry usually remains
the same, only the lattice parameters change (distortion), leaving
weak traces on electron scattering. Secondly, the small change in
resistivity can be easily covered up by effects such as
percolation, and is undetectable unless the crystal is of very
high quality. Therefore, in the present work on pure single
crystalline Bi-2212, it is interesting that we observed unusual
features in the normal-state resistivity in temperature regions
where x-ray and mechanical studies had revealed anomalous
structural properties. This is a clear indication for a
correlation between these features probed by different methods.
Moreover, drastically different temperature dependence of
resistivity appeared below these anomalies, showing crucial
connections between the phenomena in these temperature regimes.

   In our extensive resistivity measurements of both Bi-2212 and
Bi-2223 whiskers, we often noticed the data were fluctuating in a
temperature region between 180 and 270 K. Outside this window, the
signal is much more stable. Since this temperature range
corresponds to the important regime where the resistivity shows
dramatic changes and is central in the normal-state theory, we
were motivated to check the region more closely. In this work we
report five samples selected from many others which exhibit
anomalous behavior in this temperature window.

   Samples of Bi-2212 whiskers were grown by a sintering method
\cite{13}. They are rectangular needle-like tapes with typical
dimensions: 3 mm $\times$ 10 $\mu$m $\times$ 0.5 $\mu$m. The three
crystalline axes ($a$, $b$ and $c$) are along these directions,
respectively. Various characterization by scanning and
transmission electron microscopy, and energy dispersive x-ray
analysis  indicate that the whiskers are good single crystals with
perfect surface morphology throughout the whole length ($\sim$3
mm). The as-grown whiskers are usually oxygen overdoped. Before
making the first measurements, some samples were annealed in
flowing nitrogen, making them slightly underdoped. The resistivity
was measured by the standard four-wire method. The electrical
leads were made by magnetron sputtering of silver through a mask.
A dc current (0.5 $\mu$ A) was driven along the $a$-axis and
voltage was also measured in this direction.

   Figure 1a shows the resistance ($R$)data of one whisker (sample
 A, as-grown), together with a segment of $R$ above 200 K. At high
 temperature, $R-T$ is well fitted to a linear line (fitting range:
260 - 308 K) with $R(\Omega) = 129.62 + 2.475T$. As the
temperature is further lowered, $R$ shows a clear upward curving
just before it follows the well-known downward deviation. To
examine the curve in more detail, we plotted the temperature
dependence of the resistance deviation $\Delta R \equiv R-(a +
bT)$, where $(a + bT)$ is the high temperature fit as described.
The sharp turning point and the slope change are more evident in
the plot of $\Delta R$ as shown in Fig. 1b. We can see that, while
$R$ keeps decreasing with $T$, $\Delta R $ starts to deviate {\it
upward} at $T_0 \sim 260$ K and reaches a maximum at $T^{*} =236$
K, it then falls and becomes negative, i.e. $R < a + bT$. The peak
$\Delta R/R$(300 K) is about 0.5\%. Although the choice of fitting
range of temperature may slightly affect the kink, the two major
features that $R$ curves upward from linear-$T$ dependence and
reaches a sharp turning point are clearly identifiable. Such small
bumps could be easily smeared out in samples with defects such as
grain boundaries, where percolation is not negligible, to yield a
fully linear-$T$ resistance. In our twined samples, which have a
two-step superconducting transition, linear-$T$ dependence was
observed without any feature in the normal state. This agrees with
the studies by Gorlova and Timofeev \cite{14} who made combined
electron diffraction and resistivity measurement on the same
sample, and showed that a wider portion of linear-$T$ resistivity
appears in twinned whiskers or samples covered with a
polycrystalline film than in purer samples. This explains the
often-observed discrepancy in bulk materials between transport and
other measurements in which the sample's polycrystalline nature is
not important. The smallness in the change of resistivity puts
high requirement on sample quality to observe these anomalies. The
features were identified only after more significant kinks around
this temperature region were found in our Bi-2223 single crystal
samples \cite{15}. The kink in $\Delta R$ seems to follow a
power-law relation $(1-T/T_0)^{2}$ (dotted line in Fig. 1b, where
$T_0 = 260$ K). However, it could also fit to a weak exponential
increase of the form: $R_{0} = \exp[c(T_{0} - T)]$, with $c = 0.1$
to 0.2.

   In Fig.2 we plotted  $\Delta R/R$(300) vs. $T$ for four more
samples plus sample A from Fig. 1b. The annealing status of these
samples are as follows: A and B: as-grown in ambient pressure; C:
annealed in flowing N$_2$ ($450^\circ$C, 8 h); D: flowing N$_2$
annealing ($450^\circ$C, 4 h); and E: flowing O$_2$ annealing
($450^\circ$C, 10 h). The doping level may not solely depend upon
the annealing conditions because the sample size varies.
Nevertheless, data from long-time annealed samples are reliable.
Two cases in the temperature dependence of $\Delta R$ can be
classified, as shown in the separate panels of Fig. 2. In the
lower panel, $\Delta R$ reveals one kink as discussed. $\Delta R$
falls steeply right after the sharp peak, which gives rise to the
well-known downward turning $R-T$ in underdoped cuprates. Our
identification of this transition temperature is much clearer than
the data available in the literature \cite{16}. The curves in the
top panel show another kink following the first one (samples B and
C), which resulted in an S-shaped feature in the $R-T$ curve. At
the end of this curving, $R$ nearly resumes the previous
linear-$T$ descent. The origin of this anomaly is unclear. It does
not show up in the underdoped cases. For the sample annealed in
oxygen (E), the kink around 250 K is absent, but the kink at lower
temperature ($\sim$ 220 K) is consistent with the second kinks in
the other two samples. A very similar single kink was also found
in the as-grown data of sample C (refer to the inset of Fig. 3),
which is also oxygen overdoped. These features in the normal state
thus seem to be intrinsic to the Bi-2212 samples. This result
joins our previous results from Bi-2223 samples to give a
consistent picture about the anomaly around this characteristic
temperature region in Bi-series cuprates. The peaking temperature
$T^{*}$ shows a trend to increase with the extent of underdoping.
$T^{*}= 254$ K is the highest for sample C which experienced the
longest N$_2$ annealing, whereas the oxygen-annealed sample E
showed the lowest $T^{*}$ (220 K). The temperature $T_0$, at which
$\Delta R$ starts to climb up, also shows the similar dependence.
The detailed characteristic temperatures for the samples are
listed in Table I. In light of the thermal and elastic measurement
results in the literature, which showed evidence of structural
phase transformation around the same temperature region as is
reported here, we interpret such upward deviation as due to
structural transformation or lattice distortion. Similar kinks in
resistivity had been observed in La-214 due to structural
transformation \cite{12}. The signal is often fluctuating
immediately following the kink, in the region around 200 K as
shown by the oscillating $\Delta R$. This is obviously correlated
to the $\Delta R$ jump, and may well mean an influence on electron
scattering due to structural instabilities.

    The peak temperature $T^*$ clearly separates two regions in which
different mechanisms are switched on. Especially for the noisy
region below $T^*$, if it is caused by the opening of a pseudogap
in the spin excitation, the initial fluctuating $R$ data may be a
clue for this mechanism, when the antiferromagnetic correlation is
gradually frozen out. In the scenario of stripe phase formation,
on the other hand, the stripe phase may be triggered by the
lattice distortion. Charge or spin excitation following a
structure transformation (or lattice distortion) was established
by neutron scattering studies \cite{17}. Bianconi {\it et al.}
\cite{18} have shown for Bi-2212 that 1D modulation of the CuO$_2$
plane is formed by alternating stripes of low temperature
orthorhombic and low temperature tetragonal lattices. However, our
$T^{*} (\sim 3T_c)$ is higher than their data ($\sim 1.4T_c$).
Whatever the model will be, the present result shows that the
lattice structure undergoes some change in all the samples, which
initiates new resistive mechanisms. However, the subsequent
dynamics depends on the oxygen doping level, not on this
structural change.

The upward deviation is not due to thermal stress in the sample,
which is anchored onto the glass substrate by silver leads. We
estimate the thermal stress to be a small value $\sim$ 0.018 GPa
at 250 K, by taking the Young's modulus $Y = 20$ GPa \cite{9}, the
thermal expansion coefficient along the $a$-axis $\alpha = 14.4
\times 10^{-6}$ K$^{-1}$ \cite{19}, and a length between voltage
leads of about 0.3 mm at room temperature. Chen {\it et al.}
\cite{20} had measured the effect of uniaxial stress (strain) on
the resistivity of similar whiskers, and found that the influence
is very small ($\sim 1/12000$ at room temperature). Moreover, the
pressure generally causes a decrease in resistivity.

    To further study the influence of doping on $T^*$, we annealed one
sample (C) successively in flowing nitrogen. The resistance data
from this sample are as shown in Fig. 3. The result from the
as-grown sample shows a slope change near 200 K, $R-T$ is linear
above this temperature. After one annealing, a new feature
appeared around $T^* \sim 220$ K, while the slope change around
200 K remains. After the second annealing, $T^*$ shifted to 245 K,
as indicated by the arrows, showing the tendency that $T^*$
increases with further underdoping, which agrees with the widely
observed behavior of resistivity anomaly and the pseudogap.

    All the above discussion is based upon crystal structure
instability in the normal state. In light of the magnetic
structure of the CuO$_2$ planes, an alternative interpretation may
come from the competition between quasiparticle scattering and the
increasing correlation with antiferromagnetic (AF) ordering as the
temperature is decreasing.  It is interesting to compare the
present results with the anomalous resistivity of magnetic
materials \cite{21,22,23,24}. In fact, resistivity kinks were
commonly found near magnetic phase transitions in the latter. For
example, in dysprosium single crystals \cite{21}, a kink in the
linearly decreasing $\rho-T$ appears near the N\'{e}el point when
the sample transforms from the paramagnetic into the AF state.
Similar anomalies were also reported in europium chalcogenides
\cite{22}. It is surprising that the N\'{e}el temperature of
perovskite KNiF$_3$ is about 253 K \cite{24}, very close to the
anomalous temperatures in the present work. This may point to a
common origin of the resistivity kinks for the materials with
similar crystal structures.

 In the scenario of free electron scattering
with spin excitations, the total resistivity consists of three
additive terms: the $T$-independent residual resistivity $\rho_0$
due to elastic impurity scattering, the lattice contribution
(phonon) $\rho_L \propto T$, and the spin-disorder resistivity
$\rho_M$. The resistivity kink (exponentially increasing with
decreasing $T$ ) is associated with the spin-disorder resistivity
($\rho_M$). In the critical region of magnetic transition, the
behavior of $\rho_M$ reflects the strongly temperature dependent
short-range spin-spin correlation. This is in the regime of
spin-disorder due to the large long-range magnetic fluctuations
near the transition temperature \cite{24}. The resistivity thus
takes the form:

\begin{equation}
\rho(T)=\rho_0-b|t-1|\ln|t-1|^{-1},
\end{equation}
where $t = T/T_c$. For $b>0$, $\rho(T)$ shows a concave cusp at
the magnetic transition temperature $T_c$. If this idea is valid,
then the observed anomaly near $T^* \sim 250$ K must come from the
antiferromagnetic ordering in the CuO$_2$ planes. Especially, the
sharp turning point in all the curves may well signify a similar
magnetic phase transition. Currently only the spin fluctuation
theory \cite{25,26} is a close approach to this picture.

    In conclusion, an upward deviation from the linear-$T$ dependence
were observed around $T^* \sim 250$ K in the normal-state
resistivity of Bi-2212 single crystals. This is usually followed
by a fluctuating region near 200 K. We interpret such behavior as
due to structural instabilities in accordance with many results
from other measurement techniques (thermodynamic, elastic,
acoustic, and x-ray). A peak in the resistance deviation ($\Delta
R$) is clearly identifiable. Below this sharp turning point, the
resistivity drops more steeply, signifying a new mechanism in
underdoped samples. This turning in resistivity had been widely
identified as pseudogap opening. However, the scenario of stripe
formation preceded with lattice instabilities was also proposed
\cite{17}. The result shows that the lattice instability exists in
all the samples regardless of their doping level. The same result
was also found in our Bi-2223 samples. This gives a consistent
picture intrinsic to similar crystal structures of Bi-cuprates.
Although the structural transformation seems to trigger new
mechanism, it is the oxygen doping that controls the subsequent
behavior of $R-T$. It is noticeable that our $T^*$ data are
consistently higher than the literature values (typically 120 -165
K, with the highest for Bi-2212 being 220 K \cite{16}). There is
no identifiable feature (smooth $dR/dT$) in our $R-T$ curves below
180 K. An alternative interpretation of the resistance kinks is
the correlation between electrons and AF spin fluctuations.
Similar mechanisms in magnetic materials are known to give rise to
the kinks in the resistivity curves at magnetic phase transition.
Our work should stimulate theoretical interest in this direction.

\begin{figure}
\caption{(a) Resistance of one Bi-2212 whisker with $T_{c}$ = 86.6
K (obtained from the  maximum $dR/dT$). The lower curve is a
detailed view for $T =$ 200 to 260 K. The jump in slope is evident
near 236 K as indicated by the arrow. The solid line is a
linear-$T$ fit at high temperature (250 to 310 K); (b) Resistance
deviation, $\Delta R \equiv R - (a + bT)$, where $(a + bT)$ is the
linear-$T$ fit in (a). The dotted line is a power-law fit:
$\propto (260 - T)^{2}$.} \label{1}
\end{figure}

\begin{figure}
\caption{Temperature dependence of  $\Delta R$ as defined in Fig.
1b. The annealing status of the samples: A and B are as-grown; C
and D were annealed in N$_2$ ($450^\circ$C) for 8 and 4 h,
respectively; and E was O$_2$-annealed ($450^\circ$C, 10 h). Note
the noisy signal just below the kinks.} \label{2}
\end{figure}

\begin{figure}
\caption{Influence of repeated thermal annealing on the resistance
of one sample (C). Inset: $\Delta R/R$(300 K) vs. $T$. For
clarity, some fluctuating data points just below the kinks near
200 K were taken off, refer to the raw curve of open triangles.
The arrows indicate the new kinks at increasing $T^*$ after
successive N$_2$-annealing.} \label{3}
\end{figure}

\narrowtext
\begin{table}
\caption{Characteristic temperatures of the samples: $T_c$ (from
the maximum $dR/dT$), $T^*$ at the  $\Delta R$ peaks, and $T_0$
where $\Delta R$ starts to deviate upward. Also shown are the
thermal annealing conditions (always at $450^\circ$C).}
\begin{tabular}{cccc}

Samples& $T_c$ (K) & $T^*$ (K)&$T_0$ (K)\\ \tableline
A
(as-grown)&87.5 & 236& 263\\
B (as-grown)& 85 & 252& 275\\ C
(N$_2$/8h) &76& 254& 278\\
D (N$_2$/4h)&70& 245 &262\\
E (O$_2$/10
h)& 76 & 220 & 246\\
\end{tabular}
\end{table}
\widetext


\begin {references}
\bibitem{1} For a review on ultrasonic measurements, see J. Dominec,
Supercond. Sci. Technol. 6 (1993) 153.
\bibitem{2} A. Migliori, W.M. Visscher, S. Wong, S. E. Brown, I. Tanaka, H.
Kojima, P. B. Allen, Phys. Rev. Lett. 64 (1990) 2458.
\bibitem{3} J. Thoulous, X. M.Wang, D. J. L. Hong, Phys. Rev.
B 38 (1988) 7077.
\bibitem{4} S. Hoen, L, C. Bourne, C. M. Kim, A. Zettle, Phys. Rev. B
38 (1988) 11949.
\bibitem{5} X. D. Shi {\it et al.}, Phys. Rev. B
39 (1989) 827.
\bibitem{6} J. Dominec, C. Laermans, V.
Plech\'{e}\v{c}ek, Physica C 171 (1991) 373.
\bibitem{7} Ye-Ning Wang {\it et al.}, Phys. Rev. B 41 (1990) 8981.
\bibitem{8} O-M Nes {\it et al.}, Supercond. Sci. Technol. 4 (1991) S388.
\bibitem{9} T. M. Tritt, M. Marone, A. C. Ehrlich, M. J. Skove, D. J.
Gillespie, R. L. Jacobsen, G. X. Tessema, J. P. Franck, J. Jung,
Phys. Rev. Lett. 68 (1992) 2531.
\bibitem{10} Yusheng He,
Jiong Xiang, Xin Wang, Aisheng He, Jinchang Zhang,  Panggao Chang,
Phys. Rev. B 40 (1989) 7384; T. Fukami, A. A. A. Youssef, Y.
Horie, S Mase, Physica C 161 (1989) 34; V. Plech\'{e}\v{c}ek, J.
Dominec, Solid State Commun. 74 (1990) 6339.
\bibitem{11} S. Titova, I. Bryntse, J. Irvine, B. Mitchell, V.
Balakirev, J supercond. 11 (1998) 471.
\bibitem{12} Y. Nakamura, S. Uchida, Phys. Rev. B 46  (1992) 5841.
\bibitem{13} J. Jung, J. P. Franck, D. F. Mitchell, H. Claus, Physica C
 156 (1988) 494.
\bibitem{14} I. G. Gorlova, V. N. Timofeev,
Physica C 255 (1995) 131.
\bibitem{15} Weimin Chen, J. P. Franck, J. Jung, Phys. Rev. B
60 (1999) 3527.
\bibitem{16} T. Watanabe, T. Fujii, A. Mastuda, Phys. Rev. Lett. 79
(1997) 2113.
\bibitem{17} J. M. Tranquada, J. D. Axe, N. Ichikawa, Y. Nakamura, S.
Uchida, B. Nachuni, Phys. Rev. B 54 (1996) 7489.
\bibitem{18} A. Bianconi {\it et al.}, Europhys. Lett. 31 (1995) 411; A.
Bianconi, Physica C 235-240 (1994) 269.
\bibitem{19} R. H. Ardent, M. F. Garbauskas, C. A. Meyer, F. J. Rotella, J.
D. Jorgensen, R. L. Hitterman, Physica C 182 (1991) 73.
\bibitem{20} Xin-Fen Chen, G. X. Tessema, M. J. Skove, Physica C 181
(1991) 340.
\bibitem{21}P. M. Hall, S. Legvold, F. H. Spedding, Phys. Rev. 117  (1960) 971.
\bibitem{22} S. von Molnar, IBM J. Res. Develop.
 14 (1970) 269.
\bibitem{23} S. Takada, Prog. Theor. Phys. 46
(1971) 15.
\bibitem{24}P. G. de Gennes, J. Friedel, J. Phys.
Chem. Solids 4 (1958) 71.
\bibitem{25}D. Pines, in {\it High
Temperature Superconductors and the C$_{60}$ System}, ed. H-C.
Ren, (Gordon \& Breach, 1995), p.1.
\bibitem{26}Branko P. Stojkovic, D.
Pines, Phys. Rev. B 55 (1997) 8576, and references therein.
\end{references}
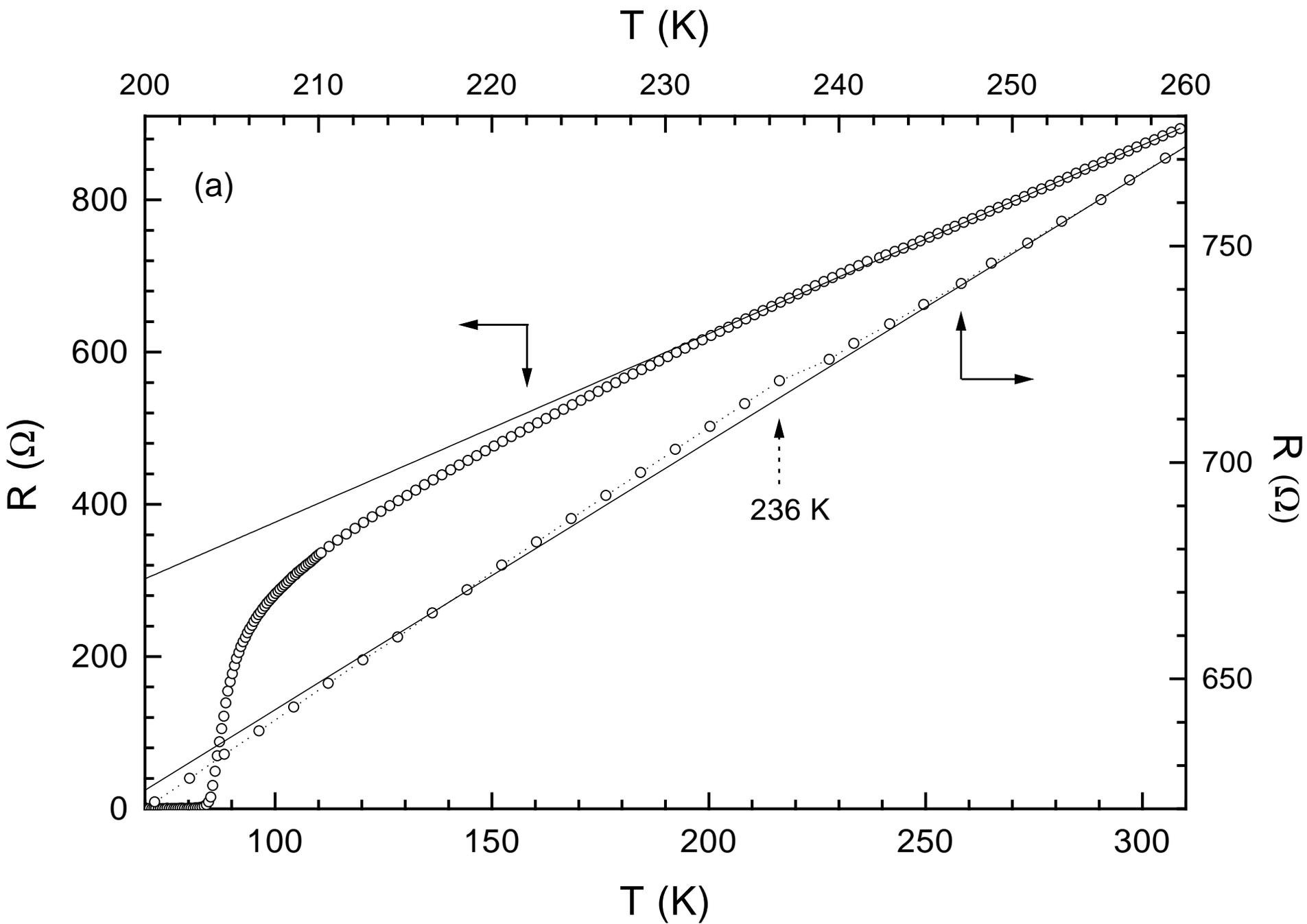
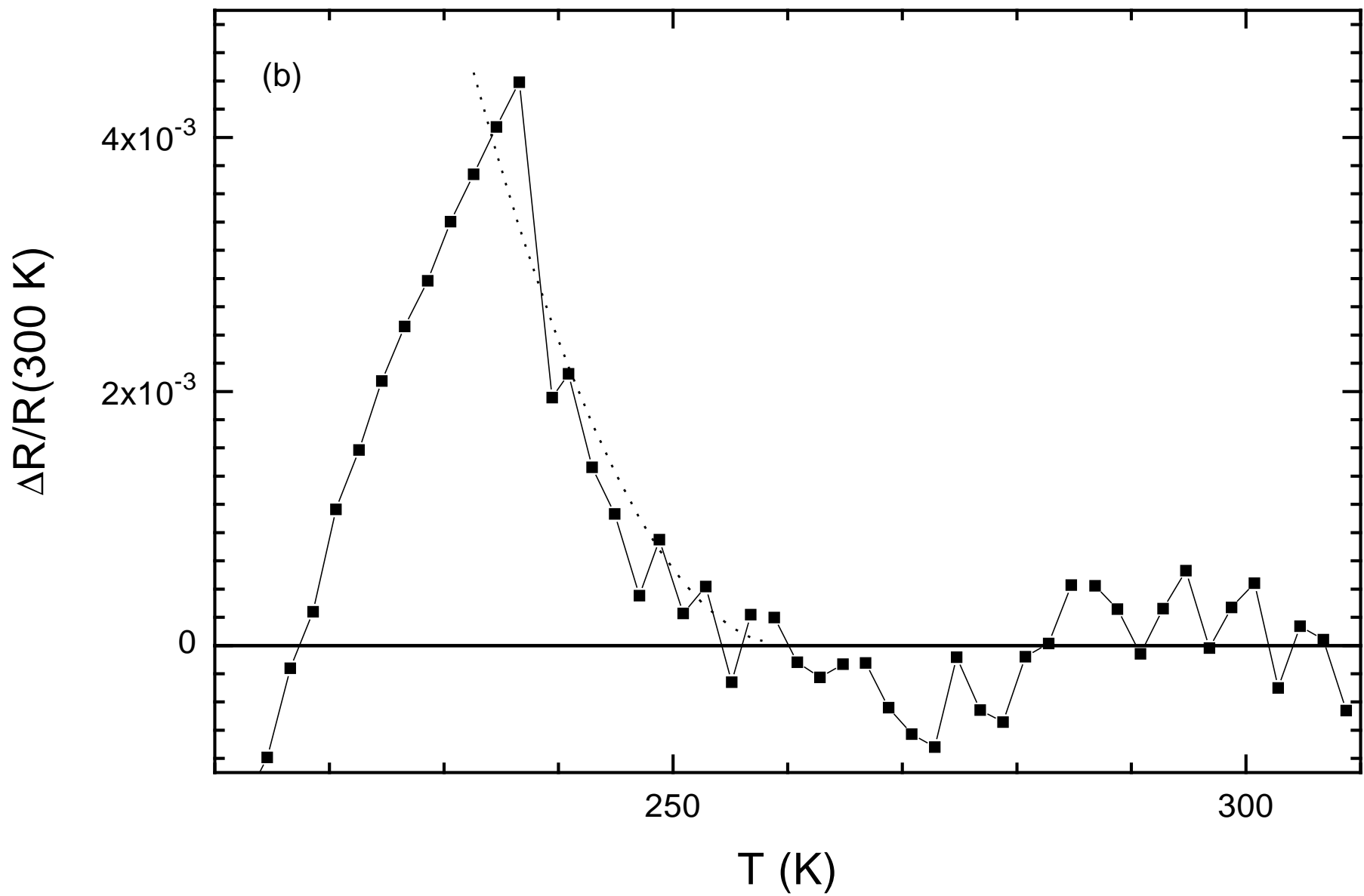
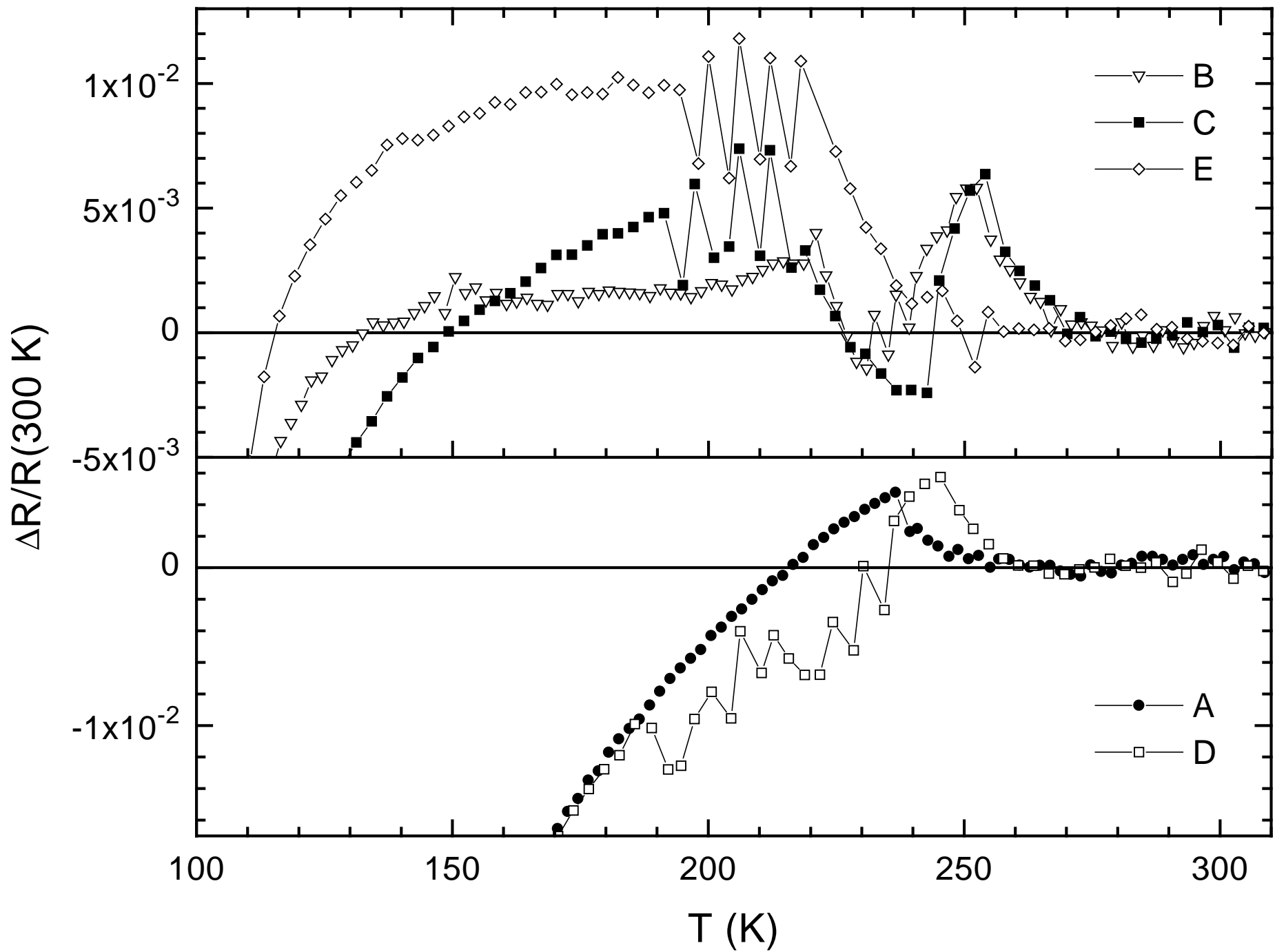
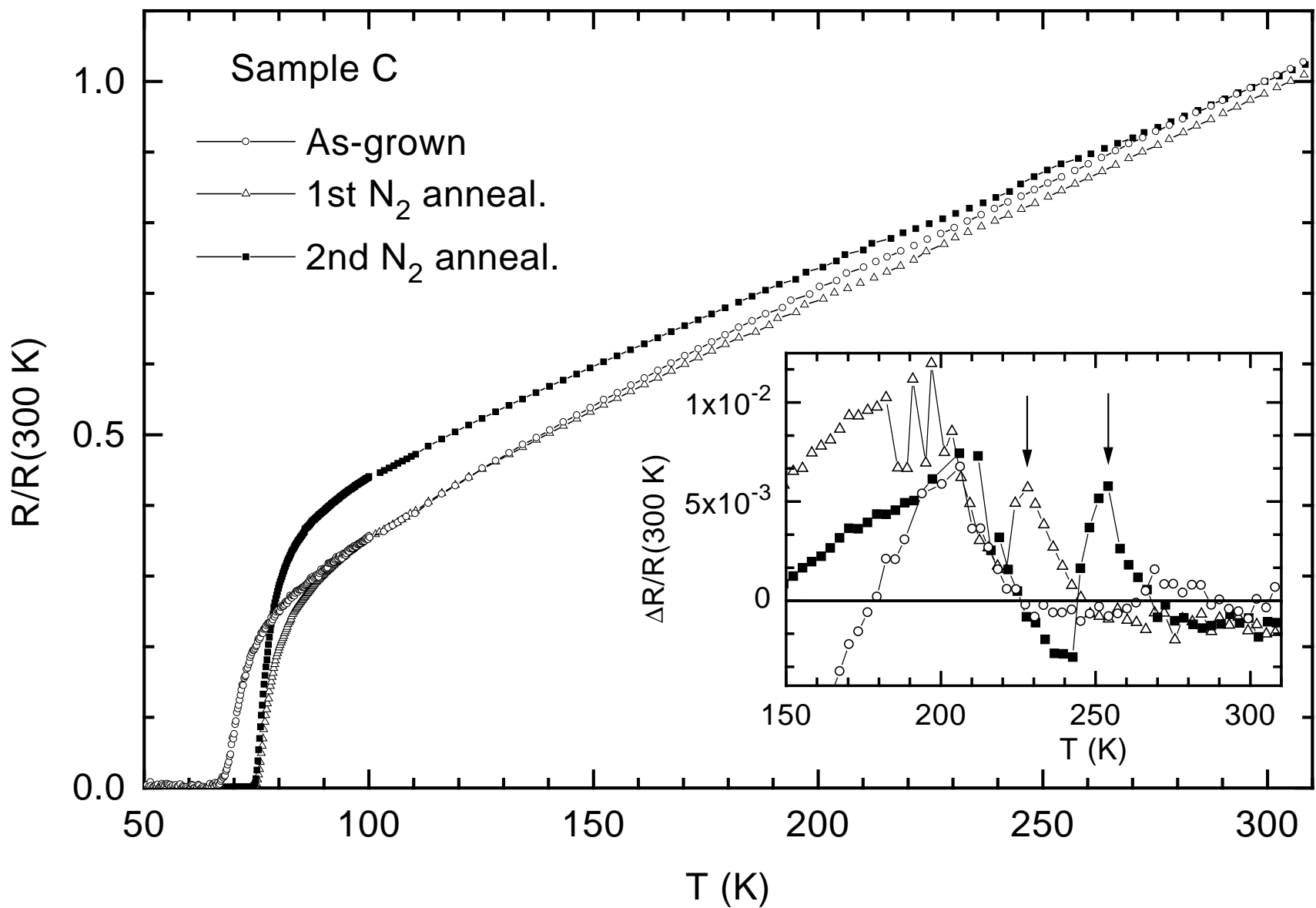
\end{document}